\documentclass[
 reprint,
 amsmath,amssymb,
 aps,
 prx
floatfix,
]{revtex4-2}
\usepackage{amsmath}
\usepackage{amsthm}
\usepackage{algorithm2e}
\usepackage{tikz}
\usetikzlibrary{shapes.geometric, arrows}

\usepackage{array}
\usepackage[caption=false]{subfig}
\usepackage{chemfig}

\usepackage{mathtools}
\usepackage{float}

\DeclarePairedDelimiterX{\norm}[1]{\lVert}{\rVert}{#1}

\usepackage{bbold}

\usepackage{mathrsfs}

\usepackage{bm}

\usepackage{physics}

\usepackage{enumitem}

\usepackage{xcolor}
\usepackage{graphicx}
\usepackage{dcolumn}
\usepackage{bm}

\begin{document}

\preprint{APS/123-QED}

\title{The Prevalence of Straight Shots in Control Space for Reaching Optimal {Quantum} Performance }
\author{Yiyou Chen}%
\email{yiyou.chen@princeton.edu}
\thanks{Yiyou Chen and Gabriele Herr contributed equally to this work. }
\affiliation{Department of Electrical and Computer Engineering, Princeton University, Princeton, NJ 08544, USA}
\author{Gabriele Herr}
\email{gh175@duke.edu}
 \affiliation{Bergen County Academies, 200 Hackensack Ave, Hackensack, NJ 07601, USA}
\author{Herschel Rabitz}
\email{hrabitz@princeton.edu}
\affiliation{
 Department of Chemistry, Princeton University, Princeton, NJ 08544, USA
}%
\date{\today}

\begin{abstract}
Gradient-based algorithms {are often} applied to identify {an} optimal control for {maximizing a} {quantum} objective function{al} {by following a steepest ascent path to the top of the control landscape}. {In contrast, this paper explores straight shots through control space where movement along such a path corresponds to a {less demanding} monotonic {complete} climb up the landscape for the {optimal} objective value.} Existing studies {of} control landscapes have revealed the presence of straight shots, but identifying them required {computationally intensive} stochastic methods. In this paper, we introduce a non-stochastic straight-shot search algorithm (STSA) that {identifies a large} class of straight shots. We apply STSA to {numerically} generate {more than} 10,000 distinct straight-shot optimization paths and analyze their statistical distribution {for each of the objectives of (i) state-to-state transitions, (ii) general observables, and (iii) unitary transformation generation.} The results show a decrease in the nearest-neighbor distances among straight shots as the sample size increases, suggesting their prevalence in control space. We further conjecture that the starting locations of the straight-shot pathways {on a landscape level set} may form connected neighborhoods, motivating future investigation into their underlying topological structure. {The present paper focuses on the use of STSA for locating massive numbers of straight solutions. {However, the STSA is not a high-efficiency algorithm in its present form, but the discovered prevalence of control straight shots leaves open future advances to accelerate the discovery of straight shots {as optimal controls}.}}
\end{abstract}

\maketitle


\section{Introduction}
\label{sec:intro}The ability to control quantum phenomena in atoms and molecules has various applications, including exciting a molecule to a target state, simultaneous optimization
of several chemical properties, optimal discrimination of multiple quantum systems, and mimicking the behavior of another material {as well as application in quantum information sciences} \cite{Brif_2010,rabitz2004quantum, werschnik2007quantum, judson1992teaching, peirce1988optimal, kosloff1989wavepacket, kirk2004optimal, macki2012introduction,zhu1999managing, gross1993inverse,campos2017make, mccaul2020driven, PhysRevA.78.033414,PhysRevA.77.063425, li2002optimal, 10.10631.1883170,GabrielTurinici_2004, karpinski2021control,pravia2003robust}. Quantum optimal control theory (OCT) concerns identifying controls that optimize a given objective function \cite{PhysRevLett.68.1500, PhysRevA.37.4950,10.10631.454384,KOSLOFF1989201}. With the advances in {applied field}-pulse shaping technologies, quantum control experiments have been performed {that} {support} the theory \cite{JAKUBETZ1990100, 10.10631.48216, PhysRevA.80.013408,PhysRevA.79.053417}.
\begin{figure}[htbp]
\includegraphics[width=0.95\linewidth]{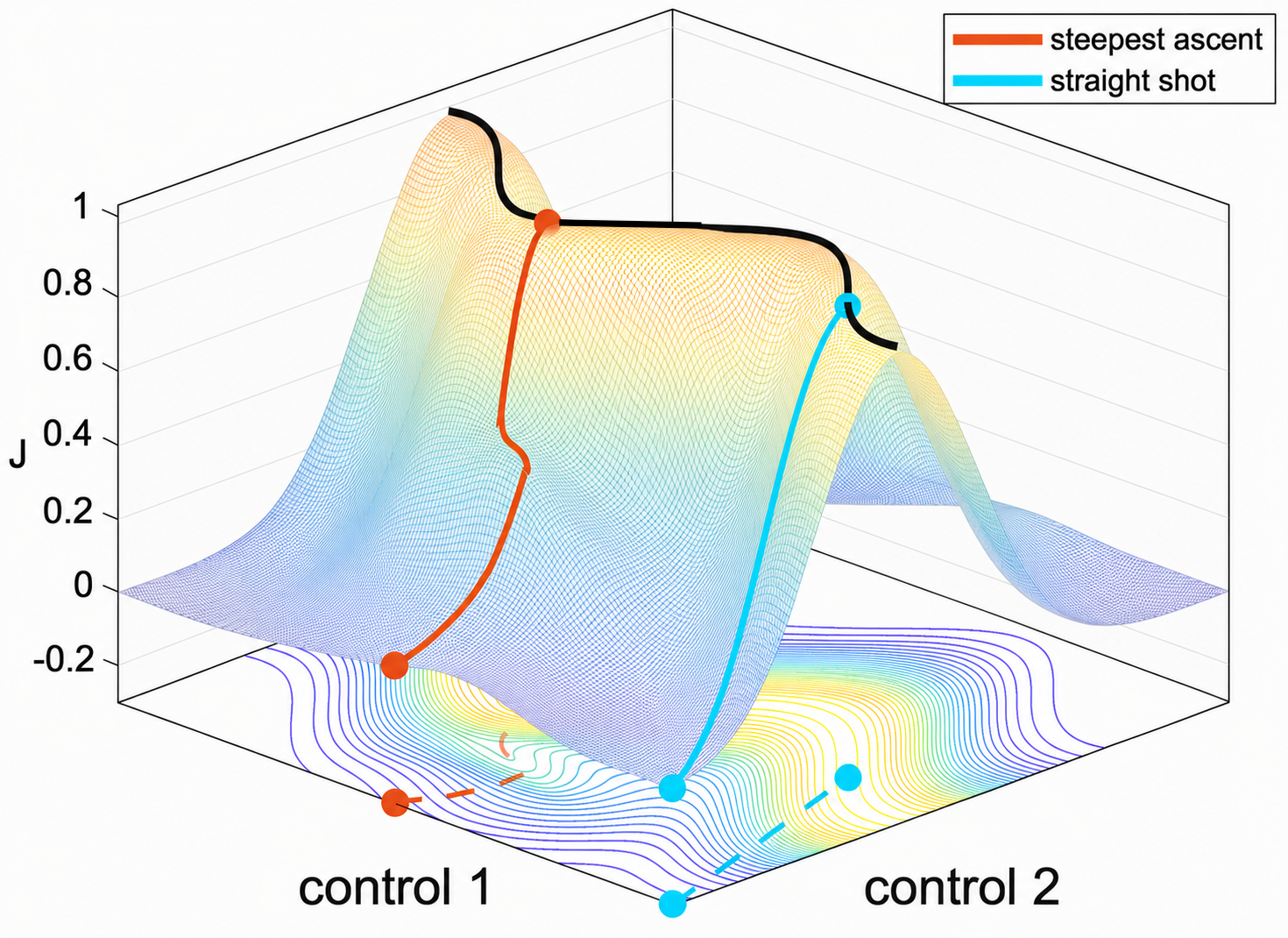}
    \caption{A schematic illustration of an optimization landscape with two {cases} representing a steepest ascent climb (red) and a straight-shot {monotonic control} climb (blue), both reaching the optimal value of the objective. {Note that the figure suggests, as has been numerically demonstrated \cite{PhysRevA.93.023427}, {that} the gradient climb from the initial {blue} dots likely would produce a monotonic optimal climb. {The schematic {shows} only two controls, but in practice there will be many.}}}
    \label{fig:cartoon}
\end{figure}

Various gradient-based algorithms have been developed to identify optimal controls, but few have investigated {the nature of the pathways through control space {taken} while seeking optimality} \cite{werschnik2007quantum,goerz2019krotov,abdelhafez2019gradient,li2017hybrid, machnes2015gradient,chen2023accelerating,li2023efficient}. In \cite{PhysRevA.88.033425}, a {metric denoted as $R$} was introduced to measure the straightness of optimization {paths} {through control space} (see Eq. (\ref{R-expression})). {This metric} motivated subsequent investigations into {{the} special class of straight {paths} through control space {whose {demonstrated} existence is intriguing} {along with the possibility of opening} up new, and ideally, efficient {search} procedures} \cite{nanduri2015exploring, PhysRevA.93.023427,PhysRevA.90.013406,PhysRevA.88.033425}. {{An} earlier} {exploration} {constructed random straight {paths} by connecting arbitrary control {pairs determined to be} {respectively} on the top and bottom level sets \cite{PhysRevA.93.023427}; surprisingly, the objective function almost {always} increased monotonically along all of these paths. In addition, steepest ascent was combined with a stochastic algorithm to identify initial controls that lead to straight control space optimization paths \cite{PhysRevA.88.033425, nanduri2015exploring}.} {Building on this prior work,} in this paper, we argue that the {optimized objective fidelity} does not necessarily {need to} follow the steepest-ascent direction {while still corresponding to a straight control {path}}. In particular, we identify a more general class of optimization paths, referred to as straight shots, which are formally defined in Section \ref{sec:optimal-st}. In Fig. \ref{fig:cartoon}, we provide a schematic illustration distinguishing straight {control} shots {with monotonic climbs} from steepest-ascent {climbs}, both reaching the maximal fidelity of the objective function. Identifying straight shots capable of achieving the maximal fidelity {provides} a further understanding of the topological structure of the optimization landscape (e.g., the inner product between any two gradients {of the objective} {associated with} straight shots stays positive). We demonstrate that such straight shots are prevalent in control space {through designing} a non-stochastic algorithm to search for them. Although this work {focuses on identifying straight control paths, the procedure involved} does not {outperform }existing optimal control algorithms. {The demonstrated wide prevalence of straight optimal control shots} motivates future research into {seeking} more effective methods for identifying optimal controls.

This paper contains two main objectives. First, a \textbf{ST}raight-shot \textbf{S}earch \textbf{A}lgorithm (STSA) {is developed to identify straight shots, avoiding cumbersome stochastic search approaches \cite{PhysRevA.88.033425, nanduri2015exploring}}. Second, we examine the distribution of optimal straight shots and show their high prevalence in control space. In Section \ref{sec:background}, we introduce three types of quantum control problems and provide the necessary physical background. In Section \ref{sec:optimal-st}, we define straight shots and relevant terminologies such as optimal controls, level sets, bottom and top controls, and bottom and top fidelities. We also briefly describe a random walk algorithm ({i.e., }as a preliminary {motivation} {for developing} STSA), {suggesting} the prevalence of optimal straight shots. The mathematical foundation and algorithmic details of STSA are laid out in Section \ref{sec:STSA}. We then apply STSA to identify $10,000$ optimal straight shots for the state-to-state transition probability problem in Section \ref{sec:numerical}. Several intriguing features of the optimal straight shots are identified, and a conjecture regarding their topological structure is proposed. 
\section{Background}~\label{sec:background}
When a closed quantum system is controlled by an {applied field (possibly an electric field from a laser source, $\epsilon(t)$)}, the dynamics of the unitary evolution operator $U(t,0)$ {is} governed by the Schrödinger equation (setting $\hbar=1$ for convenience):
\begin{align}
\label{eq:schrodinger}&i\frac{d}{dt}U(t,0) = (H_0 - \mu\epsilon(t)) U(t,0), \indent U(0,0)=I,
\end{align}
{In this work, we consider systems with dimension $N$.} In Eq. (\ref{eq:schrodinger}), $H_0$ is the control-free part of the Hamiltonian (also called the drift matrix) and $\mu$ is the dipole matrix, both of which are Hermitian. By convention, all matrices, including $U(t,0)$, $H_0$, and $\mu$, are represented in the eigenbasis of $H_0$, denoted by $\{\ket{1}, \ket{2}, \ldots, \ket{N-1}, \ket{N}\}$, which satisfies $\bra{j}\ket{k}=\delta_{jk}$ for all $1\leq j,k\leq N$, so that $H_0$ is real diagonal. We removed an additional global phase of $U(t,0)$ by setting all diagonal elements of $\mu$ to zero. A pure quantum state is defined as a superposition of the basis states, which can be represented by a state vector $\ket{\psi}\in \mathbb{C}^N$ with unit norm. A mixed state is an ensemble of pure states represented using a density matrix $\rho \in \mathbb{C}^{N\times N}$ that is Hermitian and satisfies $\Tr(\rho) = 1$. Under the time-dependent Hamiltonian defined in Eq. (\ref{eq:schrodinger}), at time $t$, an initial state $\ket{\psi_o}$ evolves into $\ket{\psi(t)} = U(t,0)\ket{\psi_o}$, and an initial density matrix $\rho_o$ evolves into $\rho(t) = U(t,0)\rho_o U^\dagger(t,0)$. Quantum observable matrices are Hermitian{,} corresponding to measurable physical quantities. The expectation value of an observable $O$ is given by $\langle \psi \vert O \vert \psi \rangle$ for a pure state $\ket{\psi}$, and by $\mathrm{Tr}(\rho O)$ for a mixed state $\rho$. For quantum control tasks, we specify a total control time $T$ and define the control $\epsilon(t)$ on $0\leq t\leq T$. Readers may refer to \cite{d2021introduction,sakurai2020modern} for more details on quantum control and dynamics. 

Eqs.~(\ref{eq:Pif})–(\ref{eq:U}) define three classes of commonly encountered quantum optimization problems {that {each} aims to {be} maximized.}
\begin{align}
    &\label{eq:Pif} J_\psi = \abs{\bra{f}\ket{\psi(T)}}^2,\\
    & \label{eq:tr(rhoO)} J_\rho = \Tr(\rho(T) O),\\
    & \label{eq:U} J_u = -\norm{U(T,0) - W}_\textsf{fro}^2,
\end{align}
The three optimization problems defined in Eqs. (\ref{eq:Pif})--(\ref{eq:U}) can be interpreted as seeking controls whose corresponding unitary matrices $U(T,0)$ {that either} maximize the state-to-state transition probability $\mathcal{P}_{i\to f}$, the expectation value of an observable $O$, {or} the similarity {of} {$U(T,0)$} to a target unitary operator $W$, respectively \cite{Chakrabarti_2007, doi:10.1126science.1093649,10.10631.2981796,PhysRevA.83.012326,PhysRevA.81.062352,PhysRevA.72.052337}. The gradient expressions for all three quantum control problems defined in Eqs. (\ref{eq:Pif})-(\ref{eq:U}) have been derived in earlier studies \cite{Brif_2010, Chakrabarti_2007}, and are given in Eqs. (\ref{grad-psi})-(\ref{grad-U}): 
\begin{align}
&~\label{grad-psi}\frac{\delta J_\psi}{\delta \epsilon(t) } = -2\Im \bra{\psi(T)}\ket{f}\bra{f}U(T,t)\mu \ket{\psi(t)},\\
&~\label{grad-rho}\frac{\delta J_\rho}{\delta \epsilon(t) } = -2\Im \Tr\left(\rho(T) O U(T,t)\mu U(t,T)\right), \\
&~\label{grad-U}\frac{\delta J_U}{\delta \epsilon(t)} = -2\Im \Tr\left(W^\dagger U(T,t) \mu U(t,0) \right).
\end{align}
The unitary propagator $U(t, t')$ for $t' < t$ {satisfies} $U(t, t') = U(t,0)U^\dagger(t',0)$ {and is} obtained by solving the Schrödinger equation~(\ref{eq:schrodinger}) as the time-ordered exponential
\begin{align}
\label{eq:time-order-int}
U(t, t') = \mathcal{T} \exp\left( -i \int_{t'}^{t} \left( H_0 - \mu \epsilon(\tau) \right) d\tau \right).
\end{align}
In practice, we discretize the total time $T$ into {small} steps of length $\Delta t$ and approximate the control as constant within each time step. The Schrödinger equation then becomes locally time-independent.
Specifically, let $j$ and $k$ be two non-negative integers satisfying $j \leq k$. Then $U(k\Delta t, j\Delta t)$ can be approximated by
\begin{align}
~\label{eq:solve-schrodinger}&U(k\Delta t, j\Delta t)\nonumber \\&\approx e^{-i(H_0 - \mu \epsilon((k-1)\Delta t))\Delta t} e^{-i(H_0 - \mu \epsilon((k-2)\Delta t))\Delta t}\ldots\nonumber\\
    & e^{-i(H_0 - \mu \epsilon((j+1)\Delta t))\Delta t}e^{-i(H_0 - \mu \epsilon(j\Delta t))\Delta t}
\end{align}
and can be {utilized} numerically to evaluate the gradient expressions in Eqs. (\ref{grad-psi})-(\ref{grad-U}). 

The quantum control landscape properties, including the locations of the critical points (i.e., $\frac{\delta J}{\delta \epsilon(t)} = 0$ for all $t\in [0,T]$) and the eigenstructure of the Hessians (i.e., $\frac{\delta^2 J}{\delta \epsilon(t)\delta \epsilon(t')}$ for all $t, t'\in [0, T]$), of all three optimization tasks defined in Eqs. (\ref{eq:Pif})-(\ref{eq:U}) have been extensively studied. {Under satisfaction of the assumptions of } controllability, transversality, and {unrestricted} control resources, the landscapes for all three problems have been proven to be {almost always} free of traps. There are no saddles for the $\mathcal{P}_{i\to f}$ problem, and all saddles have zero volume for the $\Tr(\rho O)$ and unitary problems \cite{10.10631.2981796, PhysRevA.72.052337,
Chakrabarti_2007,ho2006effective}. These landscape properties enable simple gradient-based optimization algorithms to efficiently identify optimal controls. For instance, a steepest ascent algorithm can be applied to maximize an objective function{al} $J$ by updating the control $\epsilon(t)$ according to local gradients $\frac{\delta J}{\delta \epsilon(t)}$, which leads to solving {for $\epsilon(s,t)$ along with the Schrödinger equation}, {where the parameter $s\geq 0$ labels the progress of the optimization},
\begin{align}
    ~\label{eq:oct_grad}&\frac{\partial \epsilon(s, t)}{\partial s} = \eta(s)\frac{\delta J(s)}{\delta \epsilon(s,t)},
\end{align}
where $\eta(s)>0$ controls the rate of {the control field} update \cite{werschnik2007quantum}. The algorithmic choice made for $\frac{\partial \epsilon(s,t)}{\partial s}$ in Eq. (\ref{eq:oct_grad}) ensures {that}
\begin{align}
    ~\label{eq:mono-req}\frac{dJ(s)}{ds} = \int_0^T \frac{\delta J(s)}{\delta \epsilon(s,t)}\frac{\partial \epsilon(s,t)}{\partial s} dt \geq 0.
\end{align}
 {Over the interval} $0\leq s\leq S$, the {sequence of} differential changes in the control, e.g., in Eq. (\ref{eq:oct_grad}), form a {path} in control space. To measure the {curved nature} of optimization pathways, a metric $R$ was introduced in \cite{PhysRevA.88.033425}, defined as the ratio of the control {excursion's} total path length to its Euclidean {distance}:
\begin{align}
    ~\label{R-expression}&R = \frac{\int_0^S  \left(\int_0^T  \left(\frac{\partial \epsilon(s,t)}{\partial s}\right)^2 dt\right)^{\frac{1}{2}}ds}{\left(\int_0^T \left(\epsilon(S,t) - \epsilon(0,t)\right)^2 dt\right)^{\frac{1}{2}}}.
\end{align}
For steepest ascent defined in Eq. (\ref{eq:oct_grad}), $R$ {{will} be} minimized {corresponding to} when the {path} of $\epsilon(s,t), 0\leq s\leq S$ is straight in control space, which is equivalent to the gradient of the control landscape satisfying {$\frac{\delta J(s)}{\delta \epsilon(s,t)}=\eta^{-1}(s) \frac{\partial \epsilon(s,t)}{\partial s} = \alpha(s)\frac{\delta J(0)}{\partial \epsilon(0,t)}$, {with} $\alpha(s)>0$ for all $0\leq s\leq S$.} In the next section \ref{sec:optimal-st}, we formally define {how to determine} such {straight} optimization pathways {without {exploring} demanding steepest-ascent climbs.}
\section{Optimal straight shots}~\label{sec:optimal-st}
We formally define \textit{straight shots} as paths in control space that satisfy the following two conditions:
\begin{align}
    ~\label{eq:monotonicity}\text{(Monotonicity) } &\forall 0\leq s\leq S: \ \frac{dJ(s)}{ds}\geq 0.\\
    ~\label{eq:straight}\text{(Straightness) } &\forall 0\leq s\leq S: \ \frac{\partial \epsilon(s,t)}{\partial s} = \left[\frac{\delta J(s)}{\delta \epsilon(s,t)}\right]_{s=0}.
\end{align}
{Note that Eq. (\ref{eq:monotonicity}) encompasses steepest ascent, but does not demand it by specifically including monotonicity.} By definition, a straight shot travels from a low-fidelity control to a high-fidelity control for $0\leq s\leq S$. For convenience, we refer to the two end points of a straight shot $\epsilon(s=0,t)$ and $\epsilon(s=S,t)$ as the \textit{bottom control} and \textit{top control}, respectively.  We further define the fidelities achieved by the bottom and top controls as the \textit{bottom fidelity}, denoted by $J(s=0)$, and the \textit{top fidelity}, denoted by $J(s=S)$, respectively. {Most practical} optimal control problems aim to identify controls that achieve a target fidelity sufficiently close to $\max J$, e.g., $J=0.99$ when $\max J=1$, {which are} often referred to as {achieved by} \textit{optimal controls}; we define the \textit{top level set} to be the set of all optimized controls that achieve the target fidelity. Analogously, we define \textit{optimal straight shots} to be {those} whose top control {results in reaching} the top level set {at time $T$} ({{specifically}}, whose top fidelity exceeds the target fidelity), i.e., $\epsilon(S,t)$ satisfies the following optimality condition:
\begin{align}
~\label{eq:optimality} \text{(Optimality) }&J(S) \geq \max J - \zeta.
\end{align}
where $\zeta$ is determined by the target {tolerance} (e.g., $\zeta=0.01$ when the top level set {satisfies} $J=0.99$). For simplicity, we assume a constant velocity {(i.e., defined as $\frac{\partial \epsilon(s,t)}{\partial s}$)} along the straight shot with respect to $s$ in Eq. (\ref{eq:straight}).  Note that the straightness condition in Eq. (\ref{eq:straight}) {with constant velocity} is equivalent to the following:
\begin{align}
    ~\label{eq:straight2}&\epsilon(s,t) = \epsilon(0,t) + s\left[ \frac{\delta J(s)}{\delta \epsilon(s,t)}\right]_{s=0}, \ \forall 0\leq s\leq S;
\end{align}
{{corresponding to} $\alpha(s)=1$ in Section \ref{sec:background}.}

Recently, a sufficient condition for optimal straight shots following steepest ascent {landscape climbs} was proposed \cite{PhysRevA.90.013406, nanduri2015exploring}, and a stochastic particle swarm optimization algorithm was accordingly {utilized} {to search for such straight shots} \cite{PhysRevA.88.033425}. However, these efforts only identified {a modest number} of optimal straight shots that adhere to steepest ascent defined in Eq. (\ref{eq:oct_grad}), as opposed to a {larger number} of {straight shots} satisfying the {relaxed} monotonicity condition defined in Eq. (\ref{eq:monotonicity}), which holds whenever
\begin{align}
    ~\label{eq:gen-sol}&\forall 0\leq s\leq S: \frac{\partial \epsilon(s,t)}{\partial s} = \int_0^T A(s, t, \tau) \frac{\delta J(s)}{\delta \epsilon(s,\tau)}d\tau 
\end{align}
for any positive semi-definite {matrix} $A(s,\cdot, \cdot)$, rather than the {case with steepest ascent where} $A(s,\cdot, \cdot) = \eta(s) \mathbb{1}$ {for all $s\geq 0$, as shown} in Eq. (\ref{eq:oct_grad}).
The tremendous flexibility in $A(s,\cdot, \cdot)$ motivated {the present paper} to explore optimal straight {climbs} beyond those restricted to steepest ascent.

To avoid artifacts arising from {distinct} initializations of $\epsilon(s=0,t)$ {with different fidelity values}, we constrain straight shots to have a predefined bottom fidelity, e.g., $J(s=0)=0.01$. {Thus}, the bottom controls must {produce a $J$ value} referred to {here} as {being on} the \textit{bottom level set}, {which is sufficiently close to the true bottom at $J=0$.} We adopt {the} D-MORPH {algorithm}, a commonly used level-set exploration {procedure}, to differentially move a control {that maintains $J(s=0)$} on the bottom level set \cite{rothman2005quantum, rothman2006exploring, rothman2005observable, beltrani2007photonic, dominy2008exploring}. {Introducing an additional parameter $s'\geq 0$ to characterize the control movement along the bottom level set, where $J(s',s=0)$ remains constant, we denote the starting control of a straight shot at $s'$ by $\epsilon(s',s=0,t)$. Here, $s$ parameterizes the associated climb, as defined in Eqs. (\ref{eq:monotonicity})--(\ref{eq:optimality}).} {At each $s'$ step of the D-MORPH algorithm, {we demand $\left[\frac{\partial J(s', s))}{\partial s'}\right]_{s=0}=\int_0^T \left[\frac{\delta J(s', s)}{\delta \epsilon(s', s, t)}\right]_{s=0}\frac{\partial \epsilon(s', s=0, t)}{\partial s'}dt=0$, such that the bottom fidelity $J(s', s=0)$ remains invariant with respect to $s'$.} More specifically, we evaluate the local gradient {$\left[\frac{\delta J(s',s)}{\delta \epsilon(s', s, t)}\right]_{s=0}$ and set $\frac{\partial \epsilon(s', s=0,t)}{\partial s'}$ to be the projection of a free function $f(s',t)$ onto the null space of the gradient, i.e.,}
\begin{widetext}
\begin{align}
~\label{eq:dmorph}
\text{(D-MORPH)} \quad
\frac{\partial \epsilon(s', s=0, t)}{\partial s'} 
&=
\left(
\mathbb{1} - 
\frac{\ket{\left[ \frac{\delta J(s', s)}{\delta \epsilon(s',s,t)} \right]_{s=0}}\bra{
\left[ \frac{\delta J(s', s)}{\delta \epsilon(s',s,t)} \right]_{s=0}}
}{
\bra{
\left[ \frac{\delta J(s', s)}{\delta \epsilon(s',s,t)} \right]_{s=0}}\ket{
\left[ \frac{\delta J(s', s)}{\delta \epsilon(s',s,t)} \right]_{s=0}
}
}
\right)\ket{f(s',t)}
,
\end{align}
\end{widetext}
{using state vector notation} with the free function $f(s',t)$ guiding the movement of $\epsilon(s',0,t)$ on the bottom level set; the inner product of two real functions $u(t), v(t)$ is defined as
\begin{align}
    ~\label{eq:inner-prod}&\bra{u(t)}\ket{v(t)} = \int_0^T u(\tau)v(\tau)d\tau.
\end{align} 
{As a summary of what follows, the movement on the bottom level set seeks a point $s'$ where Eq. (\ref{eq:straight2}) can be used to make a monotonic climb up the landscape to the top level set, as explained later.}} {In the numerical calculations, {the} lower and upper level sets {were} moved even closer to {their respective} limiting values {to assess stability of the STSA.}}
\subsection{Landscape climb algorithm}~\label{sec:climb-alg}
Given a control $\epsilon(s', s=0, t)$ {corresponding to $J$ being} on {the} bottom level set, we can apply Eqs. (\ref{eq:monotonicity}) and (\ref{eq:straight2}) to compute the top fidelity. Algorithmically, it suffices to choose a sufficiently small step size, $\Delta s$, and iteratively update $\epsilon(s', s, t)$ {over $s$} using the first-order Euler's method, i.e., \begin{align}~\label{eq:climb}&\epsilon( s', s+\Delta s, t) = \epsilon(s',s,t) + \left(\Delta s\right)\left[\frac{\delta J(s',s)}{{\delta \epsilon(s',s,t)}}\right]_{s=0}\end{align} until $J(s',s)$ stops increasing {with $s$}. For a given $s'$, {the conservative procedure in Eq. (\ref{eq:climb}) is used, as the behavior of the climb is not known for the current $s'$ value.} Let $S(s')$ denote the {$s=S$} value {corresponding to} when the {identified} control {giving the maximum value of {$J(s', s)\equiv J(s', S(s'))$} is reached}. The monotonicity of the straight shot guarantees that $\frac{dJ(s', s)}{ds}\geq 0$ for $0\leq s\leq S(s')$, with $\frac{dJ(s', s)}{ds}=0$ at $s=S(s')$. This procedure described by Eq. (\ref{eq:climb}) will be referred to as the \textit{climb algorithm} throughout the remainder of the paper. {Note that the {climbing} criteria {based on} Eq. (\ref{eq:climb}) could be defined at possibly {a} higher value of $J$ upon {letting $s$ further increase}. However, the present conservative choice proved to be almost always adequate {in ultimately finding {an} $s'$ value where the landscape climb is monotonic up to the maximum value of $J(s', S(s'))=0.99$.}} {Note that repeated incremental movement on the lower level set to a new $s'$ value followed by a test climb is the essence of the STSA algorithm, more fully described in Section \ref{sec:STSA}. The net result will be a straight shot {path} in field {space} and an accompanying monotonic climb {of $J$}, as illustrated by the blue dashed line and blue curve, respectively, in Fig. \ref{fig:cartoon}.} 
\subsection{Random walk algorithm}\label{sec:random-walk-alg}
Motivated by the empirical evidence in \cite{PhysRevA.93.023427}, we {explored} whether optimal straight shots satisfying the conditions defined in Eqs. (\ref{eq:monotonicity})–(\ref{eq:optimality}) are more prevalent than those {discovered} via steepest ascent {previously with a stochastic search \cite{PhysRevA.88.033425, nanduri2015exploring}}. {Thus, as} a preliminary {motivation} to {applying} the {STSA} algorithm introduced in Section \ref{sec:STSA}, we designed a simple random walk algorithm {{to assess the ease of identifying}} optimal straight shots whose bottom controls lie on the level set, with $J=0.01$. The {random walk} algorithm takes as input a control on the bottom level set, denoted by $\epsilon(s'=0, s=0, t)$ (the details of how to get such a control will be discussed in Section \ref{sec:initial}). {Then the control} {randomly walks} {over the} bottom level set {using random small variations of the free function $f(s', t)$ in Eq. (\ref{eq:dmorph})} until its corresponding top control reaches the top level set {with a monotonic climb performed with Eq. (\ref{eq:climb}).} {Following Eq. (\ref{eq:climb}) to the top level set implies that a random walk location starting at {the final level set point} $S'$ has been found corresponding to a straight path in control space.}

{{In the} random walk reference procedure, for each $s'\geq 0$, we first use the climb algorithm defined in~\ref{sec:climb-alg} to compute the top fidelity, denoted by $J(s', S(s'))$. Then a random function $f(s',t)$ is selected that is $C^\infty$ on $\mathbb{R}$, and {we} evaluate $\frac{\partial \epsilon(s',0,t)}{\partial s'}$ {following} Eq. (\ref{eq:dmorph}). Using $\Delta s'$ to denote the size for a {random step}, we construct two new bottom controls, $\epsilon_+(s'+\Delta s', 0, t)$ and $\epsilon_-(s'+\Delta s', 0, t)$, corresponding to the two possible directions of the control update, i.e., $\epsilon_{\pm}(s'+\Delta s', 0, t) = \epsilon(s', 0, t) \pm (\Delta s') \frac{\partial \epsilon(s',0,t)}{\partial s'}$ {{using Eq. (\ref{eq:dmorph}), where $\frac{\partial \epsilon(s', 0,t)}{\partial s'}$ contains the random new $f(s',t)$}}.
We then apply the climb algorithm to both bottom controls and assign $\epsilon(s'+\Delta s',0,t)$ to whichever of $\epsilon_{\pm}(s'+\Delta s', 0, t)$ yields a higher top fidelity. 
Note that in the case when a randomly chosen $f(s',t)$ fails to increase the top fidelity in either direction at some $s'$, a new $f(s',t)$ is randomly sampled, and the same procedure is repeated with the new $f(s',t)$. 

The random walk algorithm terminates--with the value of $s'$ denoted by $S'$ {(i.e., to be used in Eq. (\ref{eq:climb}))} --when the straight shot with bottom control $\epsilon(S', 0, t)$ has a top control reaching the top level set, i.e., $J(S', S(S'))\geq 0.99$. By construction, {if successful, the random walk algorithm} returns a bottom control whose corresponding straight shot {climb}, {over} {$0\leq s\leq S(S')$,} is optimal. Furthermore, {since we update $\epsilon(s', 0,t)$ such that the top fidelity at $S(s')$ of the straight shot climb increases at all $s'$,} the top fidelity increases monotonically, i.e., $\frac{\partial J(s', S(s'))}{\partial s'}> 0$ for $0\leq s'\leq S'$. 

The random walk algorithm successfully identified distinct optimal straight shots across 1000 empirical runs, each with a different initial control on the bottom level set. The physical system used for producing the {initial controls} is specified later in Section \ref{sec:implementations}. In Fig. \ref{fig:random-walk-trajectory}, we apply Principal Component Analysis (PCA) \cite{Jolliffe2002Principal} to generate a dimensionality-reduced {image of the} random walk trajectory of the bottom control (i.e., $\epsilon(s',0,t)$ for $0\leq s' \leq S'$) {shown here} for one representative run. {Due to the {random sampling {of} $f(s',t)$ in Eq. (\ref{eq:dmorph})}}, the illustrated trajectory experiences frequent abrupt changes in direction. To quantify the difficulty of finding an optimal straight shot using the random walk algorithm, we also show in Fig. \ref{fig:distance-random-walk} a histogram of the Euclidean distances traversed by the bottom controls before reaching {the beginning of} an optimal straight shot, defined as ${\norm{\epsilon(S', 0, t) - \epsilon(0, 0, t)}_2}$, where $\norm{\cdot}_2$ denotes the $L^2$-norm, i.e., $
{\norm{\epsilon(S', 0, t) - \epsilon(0, 0, t)}_2} = \left(\int_0^T \left(\epsilon(S', 0, \tau) - \epsilon(0, 0, \tau) \right)^2 d\tau\right)^{\frac{1}{2}}$.
\begin{figure}[htbp]
\includegraphics[width=0.95\linewidth]{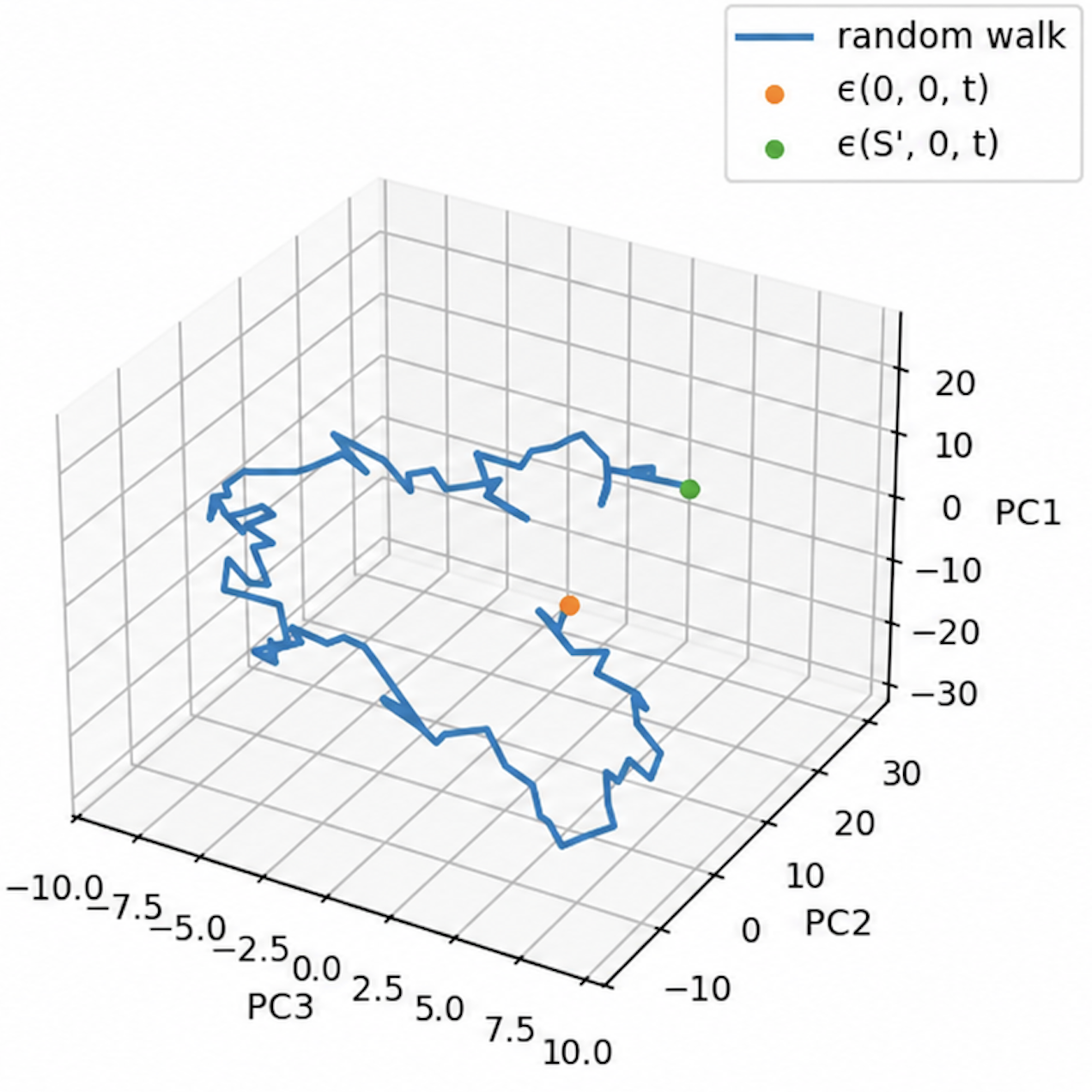}
        \caption{A dimensionality-reduced illustration of a bottom control trajectory on $J=0.01$ produced by the random walk algorithm. The controls are projected onto a three-dimensional subspace {with {principal component analysis (PCA)}} formed by the top three principal components {PC1, PC2, PC3} of all controls on the trajectory {explaining approximately 66\%, 16\%, 8\% of the total variance, respectively}. The trajectory {over the bottom level set} has an overall irregular shape with frequent abrupt changes in direction, due to random sampling of $f(s',t)$ {in Eq. (\ref{eq:dmorph})}.}
        \label{fig:random-walk-trajectory}
\end{figure}
\begin{figure}[htbp]
\includegraphics[width=0.95\linewidth]{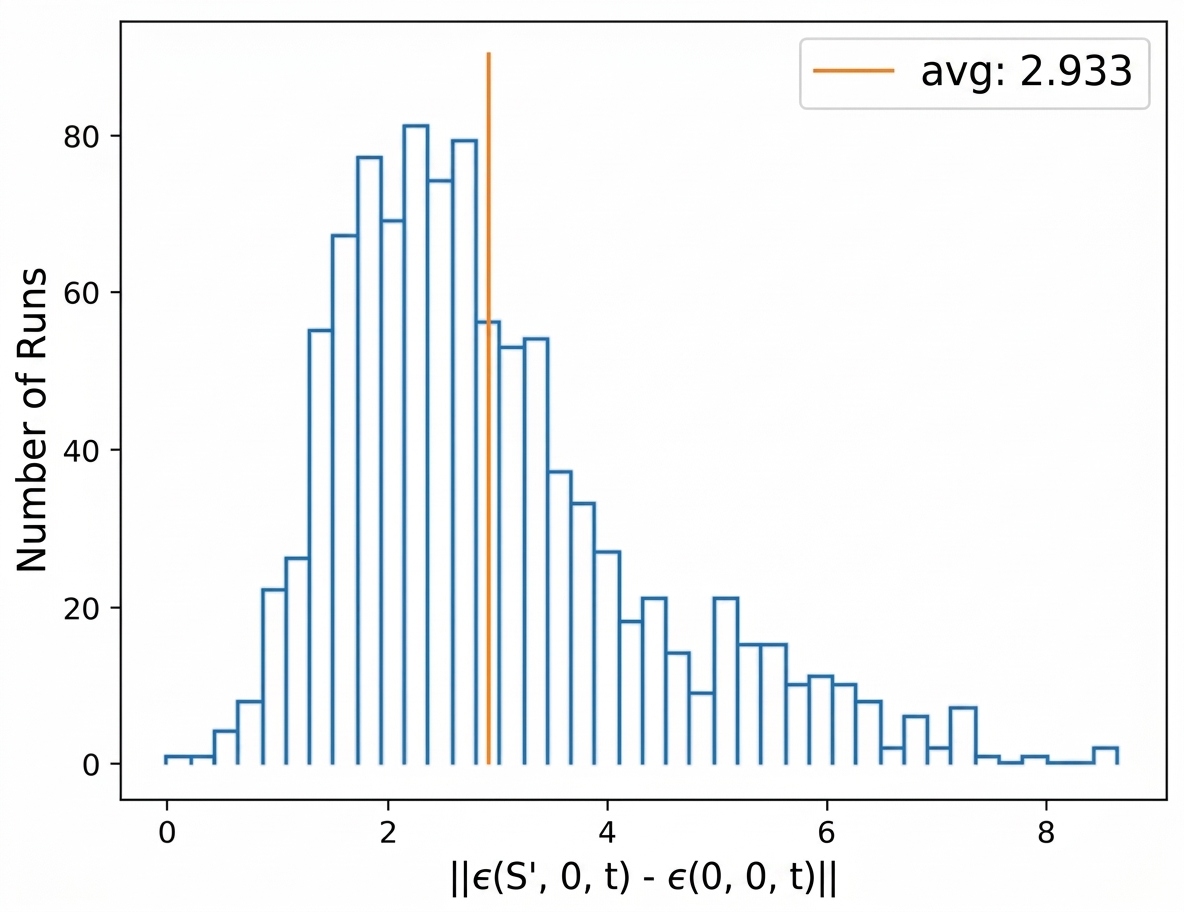}
    \caption{A histogram of the Euclidean distances traversed by the bottom controls {produced by the random walk algorithm} before finding {the initial locations $S'$ {producing} }optimal straight shots. The average distance is approximately $2.93$ across 1000 runs, which is considerably higher compared to that obtained using STSA discussed in Section \ref{sec:results}.}
    \label{fig:distance-random-walk}
\end{figure}  Under the random walk algorithm, the bottom controls traverse an average distance of $2.93$ (see Fig. \ref{fig:distance-random-walk}), which is substantially higher than the average distance of $1.31$ achieved by STSA (see Fig.~\ref{fig:distance-bottom-control}). However, {the} success in identifying optimal straight shots {via random walks} indicates their prevalence in control space, thereby motivating the development of a {non-{random walk}} search algorithm.
\section{Straight-Shot Search Algorithm (STSA)}
\label{sec:STSA}To systematically search for optimal straight shots that satisfy all three conditions defined in Eqs. (\ref{eq:monotonicity})-(\ref{eq:optimality}), we note that the top fidelity of a straight shot is completely determined by its bottom control. Namely, if we use $s'$ and $s$ to parameterize the movement of the bottom control and the straight-shot {field-driven} {monotonic} climb, respectively, then once the bottom control $\epsilon(s',0,t)$ is determined, its initial gradient $\left[\frac{\delta J(s', s)}{\delta \epsilon(s',s,t)}\right]_{s=0}$ can be evaluated, and the climb algorithm can subsequently be applied to determine the top fidelity, as discussed in \ref{sec:climb-alg}. Therefore, we consider $\epsilon(s', 0, t)$ as a functional variable of $J(s', S(s'))$ and derive a rule to update $\epsilon(s', 0, t)$ such that {the top fidelity} $J(s', S(s'))$ increases {monotonically over $0\leq s'\leq S'$, {ideally until $J(S', S(S'))=0.99$ is reached at $s'=S'$}}. By the chain rule,
\begin{align}
~\label{dJ(s',S)/ds'}&\frac{\partial J(s', S(s'))}{\partial s'} \nonumber\\&= \left[\frac{\partial J(s', s)}{\partial s'}\right]_{s=S(s')} +  \left[\frac{\partial J(s', s)}{\partial s}\right]_{s=S(s')}\frac{\partial S(s')}{\partial s'}.
\end{align}
Given the optimality condition defined in Eq. (\ref{eq:optimality}), $\left[\frac{\partial J(s', s)}{\partial s}\right]_{s=S(s')}=0$ for all straight shots {by virtue of a {monotonic} climb being taken to its maximum value}. With intermediate derivations provided in Appendix \ref{appendix1}, we obtain the expression for $\frac{\partial J(s', S(s'))}{\partial s'}$ in Eq. (\ref{eq:dj/ds'>0}).{
\begin{widetext}
\begin{align}
\frac{\partial J(s', S(s'))}{\partial s'} =&\label{eq:dj/ds'>0}\int_0^T dt \left(\left[\frac{\delta J(s', s)}{\delta \epsilon(s', s, t)}\right]_{s=S(s')} +S(s')\int_0^T dt'\left[\frac{\delta^2 J(s',s)}{\delta \epsilon(s',s,t) \delta \epsilon(s',s,t')}\right]_{s=0}\left[\frac{\delta J(s', s)}{\delta \epsilon(s', s, t')} \right]_{s=S(s')}\right)\frac{\partial \epsilon(s',0,t)}{\partial s'}
\end{align}
Since the bottom controls of the straight shots are assumed to morph on the bottom level set, we further require $\frac{\partial \epsilon(s',0,t)}{\partial s'}$ to satisfy the D-MOPRH equation in Eq. (\ref{eq:dmorph}), {restated here in Eq. (\ref{eq:de/ds'}).}
\begin{align}
\frac{\partial \epsilon(s', 0, t)}{\partial s'} =&\label{eq:de/ds'}\left(\mathbb{1} - \frac{\ket{\left[ \frac{\delta J(s', s)}{\delta \epsilon(s',s,t)}\right]_{s=0}}\bra{ \left[ \frac{\delta J(s', s)}{\delta \epsilon(s',s,t)}\right]_{s=0}}}{\bra{\left[ \frac{\delta J(s', s)}{\delta \epsilon(s',s,t)}\right]_{s=0}}\ket{\left[ \frac{\delta J(s', s)}{\delta \epsilon(s',s,t)}\right]_{s=0}}}\right)\ket{f(s',t)}
\end{align}
Substituting Eq. (\ref{eq:de/ds'}) into Eq. (\ref{eq:dj/ds'>0}), we obtain the expression for $\frac{\partial J(s', S(s'))}{\partial s'}$ in Eq. (\ref{eq:dj/ds'-dmorph}). 
\begin{align}
\frac{\partial J(s', S(s'))}{\partial s'} =&\label{eq:dj/ds'-dmorph}\bra{ \left[\frac{\delta J(s', s)}{\delta \epsilon(s', s, t)}\right]_{s=S(s')} +S(s')\int_0^T dt'\left[\frac{\delta^2 J(s',s)}{\delta \epsilon(s',s,t) \delta \epsilon(s',s,t')}\right]_{s=0}\left[\frac{\delta J(s', s)}{\delta \epsilon(s', s, t')} \right]_{s=S(s')}}\\
\nonumber &\left(\mathbb{1} - \frac{\ket{\left[ \frac{\delta J(s', s)}{\delta \epsilon(s',s,t)}\right]_{s=0}}\bra{\left[ \frac{\delta J(s', s)}{\delta \epsilon(s',s,t)}\right]_{s=0}}}{\bra{ \left[ \frac{\delta J(s', s)}{\delta \epsilon(s',s,t)}\right]_{s=0}}\ket{\left[ \frac{\delta J(s', s)}{\delta \epsilon(s',s,t)}\right]_{s=0}}}\right) \ket{f(s',t)}
\end{align}
For the top fidelity $J(s', S(s'))$ to increase monotonically with respect to $s'$, we note that $\left(\mathbb{1} - \frac{\ket{\left[ \frac{\delta J(s', s)}{\delta \epsilon(s',s,t)}\right]_{s=0}}\bra{\left[ \frac{\delta J(s', s)}{\delta \epsilon(s',s,t)}\right]_{s=0}}}{\bra{ \left[ \frac{\delta J(s', s)}{\delta \epsilon(s',s,t)}\right]_{s=0}}\ket{\left[ \frac{\delta J(s', s)}{\delta \epsilon(s',s,t)}\right]_{s=0}}}\right)$ is a positive semi-definite projector for all $s'\geq 0$. Therefore, we choose $f(s', t)$ in the form given by Eq. (\ref{eq:f}), which guarantees that $\frac{\partial J(s', S(s'))}{\partial s'}\geq 0$ under the D-MORPH algorithm.
\begin{align}
~\label{eq:f}\ket{f(s',t)} =& \ket{\left[\frac{\delta J(s', s)}{\delta \epsilon(s', s, t)}\right]_{s=S(s')} +S(s')\int_0^T dt'\left[\frac{\delta^2 J(s',s)}{\delta \epsilon(s',s,t) \delta \epsilon(s',s,t')}\right]_{s=0}\left[\frac{\delta J(s', s)}{\delta \epsilon(s', s, t')} \right]_{s=S(s')}}
\end{align}
\end{widetext}
  Consequently, we substitute Eq. (\ref{eq:f}) into Eq. (\ref{eq:de/ds'}) and obtain the gradient expression for $\frac{\partial \epsilon(s', 0, t)}{\partial s'}$.}

Note that in Eq. (\ref{dJ(s',S)/ds'}), we assumed that $J(s', S(s'))$ was differentiable with respect to $s'$. This assumption, however, may not hold when $S(s'), 0\leq s'\leq S'$ has discontinuities, potentially resulting in kinks {of} $J(s', S(s'))$ {due to the choice of stopping a climb at the first point in $s$ where $J$ can no longer locally increase, {as mentioned} in \ref{sec:climb-alg}}. {This issue, as remarked later, was a rare occurrence.} The general {analysis} developed in this section applies equally to all three quantum control problems defined in Eqs. (\ref{eq:Pif})–(\ref{eq:U}). Next, we will use the gradient expression derived in Eq. (\ref{eq:de/ds'}) to {design the} non-stochastic {STSA} search algorithm that identifies the bottom controls of optimal straight shots.
\begin{figure*}[htbp]
    \centering
    \includegraphics[width=\linewidth]{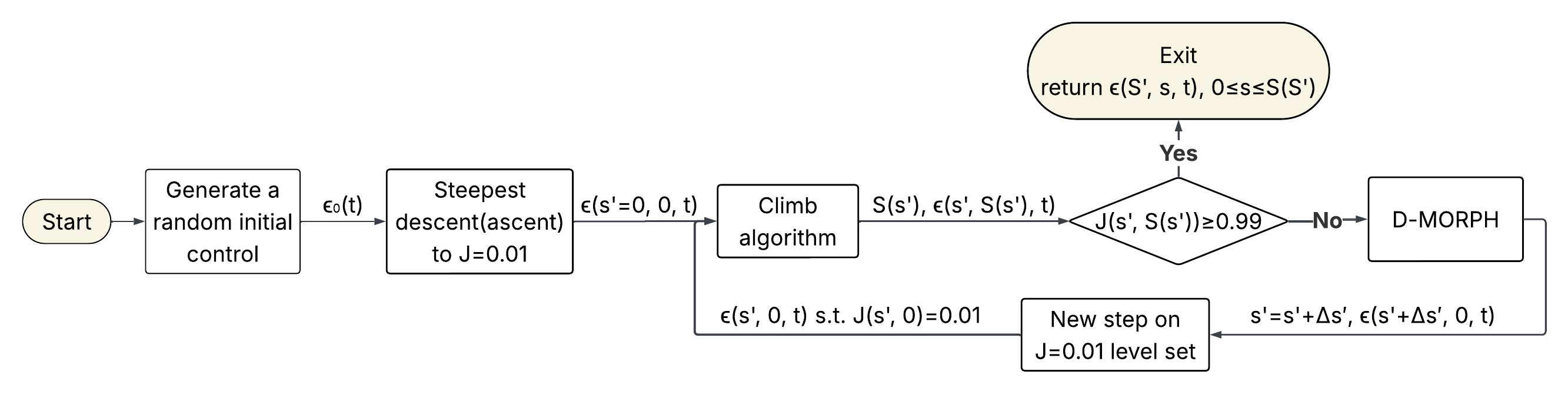}
    \caption{A flowchart showing the main components of STSA discussed in Section \ref{sec:STSA}. The initial control is randomly generated from Eq. (\ref{eq:random-controls}) and then moved down (up) to a predefined bottom level set, e.g., {to reach} $J=0.01$, using steepest descent (ascent) discussed in Section \ref{sec:initial}. STSA then iteratively morphs the {resultant} control on the bottom level set {by means of the free function $f(s',t)$ in Eq. (\ref{eq:f})} until an {initial} optimal straight {control} shot is identified {that permits monotonically climbing to the top level set, {e.g.,} $J=0.99$}. At each iteration parameterized by $s'$, a corrective step is applied to ensure that the bottom control {maintains $J=0.01$ {to an acceptable tolerance}}; the climb algorithm is then used to identify the top control (discussed in Section \ref{sec:climb-alg}), followed by D-MORPH to update the bottom control such that the top fidelity $J(s', S(s'))$ increases with $s'$ (discussed in Section~\ref{sec:STSA-alg}). }
    \label{fig:flowchart}
\end{figure*}
\subsection{Initial Control generation on the bottom level set}~\label{sec:initial}
To apply STSA, we need to first generate {an} initial control on the bottom level set. A commonly used approach is to randomly {choose a} {physically based} control and bring {the resultant initial fidelity} to the bottom level set. {{We considered} initial controls} taking the form of {Eq. (\ref{eq:random-controls})} {as} {sums of }sinusoidal functions with frequencies $\omega_{ij}$ drawn from the transition frequencies of $H_0$ {allowed by the dipole}, along with randomly chosen amplitudes $\alpha_{ij}$ and phases $\phi_{ij}$. We also normalize the {initial {trial}} control such that the maximal {control} intensity is bounded (i.e., multiplying by a global factor $\frac{1}{A}$ such that $\lvert \epsilon_o(t) \rvert \leq 1$ for all $0 \leq t \leq T$). A global Gaussian envelope is further imposed, with mean $t = \frac{T}{2}$ and standard deviation $\sigma = \frac{T}{8}$. More precisely,
\begin{align}
~\label{eq:random-controls} &\epsilon_o(t) = \frac{1}{A}e^{-{(t-T/2)^2}/{2\sigma^2}}\sum_{i < j}{\alpha_{ij}\cos(\omega_{ij} t - \phi_{ij})}\\
&\alpha_{ij}\leftarrow \text{Unif}(0,1)\\
&\phi_{ij} \leftarrow \text{Unif}[0, 2\pi)
\end{align}
{Subsequent manipulations of the field do not demand the constraints in Eq. (\ref{eq:random-controls}).}

Given a randomly generated control from Eq. (\ref{eq:random-controls}), steepest descent (ascent) is then applied to bring {the fidelity} down ({up}) to the bottom level set. Specifically, we {use $\xi$ to parameterize the descent {(ascent) to $J=0.01$} and set $\epsilon(\xi=0,t)=\epsilon_o(t)$. We then iteratively solve
\begin{align}
&~\label{eq:steepest-descent} \frac{\partial \epsilon(\xi, t)}{\partial \xi} = \beta\frac{\delta J(\xi)}{\delta \epsilon(\xi, t)}, \indent \xi \geq 0,
\end{align}
{where $\beta$ is $\pm 1$ as appropriate for ascent (descent)}}}. By the properties of the quantum control landscape discussed in Section \ref{sec:intro}, the steepest descent ({ascent}) algorithm {should be} capable of reaching a control with {the associated} fidelity arbitrarily close to the bottom level set. Let $\epsilon(\Xi, t)$ be a control obtained from steepest descent {(ascent)} that achieves a fidelity of $0.01$. We initialize the bottom control {as} $\epsilon(s'=0, s=0, t) \equiv \epsilon(\Xi, t)$ and {proceed} to solve Eq. (\ref{eq:de/ds'}).
\subsection{Identifying optimal straight shots using D-MORPH}~\label{sec:STSA-alg}
Given an initial bottom control $\epsilon(s'=0, s=0, t)$ obtained from Section~\ref{sec:initial}, we use Eqs.~(\ref{eq:de/ds'}) and (\ref{eq:f}) to {move towards} {the beginning of} an optimal straight shot. Specifically, at each $s'$ iteration with bottom control $\epsilon(s', s=0, t)$, we first apply the climb algorithm (see Section~\ref{sec:climb-alg}) to identify the top control $\epsilon(s', S(s'), t)$, and then evaluate $\frac{\partial \epsilon(s', 0,t)}{\partial s'}$ using Eq. (\ref{eq:de/ds'}). Note that the Hessian term $
\left[\frac{\delta^2 J(s',s)}{\delta \epsilon(s',s,t)\,\delta \epsilon(s',s,t')}\right]_{s=0}
$
appearing in Eq. (\ref{eq:f}) requires evaluating the second-order functional derivative of the objective function with respect to the control. Here, we {restate} the Hessian expressions for all three control problems defined in Eqs. (\ref{eq:Pif})--(\ref{eq:U}) \cite{shen2006quantum,rabitz2006topology,hsieh2009topology,shu2015hessian,shir2014efficient,xing2014assessment,rothman2006exploring}. Since the Hessians are real symmetric in those cases, it suffices to {evaluate} them for $t\leq t'$,
\begin{widetext}
\begin{align}
&~\label{eq:hessian-psi}\frac{\delta^2 J_\psi}{\delta \epsilon(t)\delta \epsilon(t')} = 2\Re \bra{\psi(t)}\mu U(t, T)\ket{f}\bra{f}U(T,t')\mu \ket{\psi(t')}- \bra{\psi(T)}\ket{f}\bra{f}U(T,t')\mu U(t', t)\mu \ket{\psi(t)}\}, \ t\leq t'\\
&~\label{eq:hessian-rho}\frac{\delta^2 J_\rho}{\delta \epsilon(t)\delta \epsilon(t')} = 2\Re \{\Tr\left(\rho_0 U(0,t)\mu U(t, T)OU(T,t')\mu U(t',0)\right)- \Tr\left(\rho_0 U(0,T)OU(T,t')\mu U(t', t)\mu U(t, 0)\right)\}, \ t\leq t'\\
&~\label{eq:hessian-U}\frac{\delta^2 J_U}{\delta \epsilon(t)\delta \epsilon(t')} = -2\Re \Tr\left(W^\dagger U(T,t')\mu U(t', t)\mu U(t, 0)\right), \ t\leq t'
\end{align}
\end{widetext}
The Hessian expressions for $t>t'$ can be obtained by swapping $t'$ and $t$ in Eqs. (\ref{eq:hessian-psi})-(\ref{eq:hessian-U}), i.e., $\frac{\delta^2 J}{\delta \epsilon(t)\delta \epsilon(t')}=\frac{\delta^2 J}{\delta \epsilon(t')\delta \epsilon(t)}$ when $t>t'$. 

With a sufficiently small $\Delta s'$, we adopt Euler's method to numerically solve Eq. (\ref{eq:de/ds'}):
\begin{align}
~\label{eq:euler-dmorph}\epsilon(s' + \Delta s', 0, t) \approx \epsilon(s', 0, t) + \left(\Delta s'\right) \frac{\partial \epsilon(s', 0, t)}{\partial s'}.
\end{align}
The STSA terminates at $s'=S'$ when the top control of the straight shot $\epsilon(S', S(S'), t)$ reaches the {first encountered} top level set, indicating that an optimal straight shot has been identified. {{As STSA is a sequential algorithm,} during the lower level set exploration at intermediate $s'$ values, the identified straight shot will correspondingly lead to an intermediate optimal yield that is $<0.99$; only at the final $S'$ value will the successful straight-shot climb, {over} $0\leq s\leq S(S')$, {and} reach $J=0.99$.}

The definition of D-MORPH in Eq. (\ref{eq:dmorph}) guarantees that the bottom control stays on the bottom level set for all $0\leq s'\leq S'$. However, numerical approximations such as time discretization and {the use of }Euler's method can introduce {small} numerical errors that cause the bottom control to drift away from the bottom level set. Therefore, a corrective step is added in each iteration to ensure that the bottom control stays close to the bottom level set. Specifically, for each $s'$, we compute the bottom fidelity, $J(s', 0)$; steepest ascent is applied when $J(s',0) < 0.01$, and steepest descent when $J(s',0) > 0.01$, until the bottom fidelity is sufficiently close to $0.01$ {(e.g., within an error of $0.001$)}. The details of the steepest ascent and descent algorithms have been provided in Eqs. (\ref{eq:oct_grad}) and (\ref{eq:steepest-descent}), respectively.

In Fig. \ref{fig:flowchart}, we present a flowchart consisting of the main components of STSA. {We note that} while STSA is guaranteed to increase the top fidelity when the differentiability assumption for $J(s', S(s'))$ in Eq. (\ref{dJ(s',S)/ds'}) is satisfied, it remains unclear whether non-differentiability at certain values of $s'$ may prevent the top control from {always} reaching the top level set. {As remarked earlier, this rarely happened.}
\section{Numerical Implementations and Results}
\label{sec:numerical} 
For illustrative purposes, we focus on the state-to-state transition probability problem in this section (see Eq.~(\ref{eq:Pif})). {The mathematical machinery in Sections \ref{sec:background} and \ref{sec:optimal-st} includes} the $\Tr(\rho O)$ and unitary {generation} problems. {Full tests were performed on these cases, and they exhibited similar qualitative behavior} ({not shown here}). 
\subsection{Implementation details}~\label{sec:implementations}
We first present the implementation details of STSA discussed in Section~\ref{sec:STSA}. For all the empirical runs, a four-level quantum system is used (i.e., $N=4$), with {arbitrarily selected} $H_0$ and the dipole matrix $\mu$ defined as 
\begin{align}
\label{eq:H_0}&H_0=\begin{pmatrix} -2.25 & 0 & 0 & 0 \\ 0&-0.75 & 0 & 0 \\ 0 & 0 & 0.25 & 0 \\ 0 & 0 & 0 & 2.75\end{pmatrix}\\
\label{eq:mu}&\mu=\begin{pmatrix} 0 & 0.75 & -0.46 & -0.10 \\ 0.75& 0 & 0.67 & -0.42 \\ -0.46 & 0.67 & 0 & -0.62 \\ -0.10 & -0.42 & -0.62 & 0\end{pmatrix}
\end{align}
For the $\mathcal{P}_{i\to f}$ problem, we {set} the initial state $\ket{i}$ to be $\ket{1}$, the lowest energy eigenstate of $H_0$, and the final state $\ket{f}$ to be $\ket{4}$, the highest energy eigenstate of $H_0$.
Furthermore, the top and bottom level sets are chosen to have fidelities $J=0.99$ and $J=0.01$, respectively. 

To ensure sufficient numerical accuracy, we set $\Delta s=0.025$ for the climb algorithm (see Section~\ref{sec:climb-alg}), $\Delta s'=0.00025$ for D-MORPH (see Eq. (\ref{eq:de/ds'})) {due to higher numerical sensitivity}, $T=20$ for the total control time, and $\Delta t = 0.05$ for time step size (see Eq. (\ref{eq:solve-schrodinger})). To keep the bottom fidelity sufficiently close to $0.01$, a Runge–Kutta–Fehlberg method with adaptive step size in $\xi$ is employed for steepest ascent (descent), both in the initialization of bottom controls (see Section \ref{sec:initial}) and in the corrective step of STSA (see Section \ref{sec:STSA-alg}) \cite{fehlberg1970classical}.
\subsection{Numerical results}
\label{sec:results}STSA is applied to 10,000 randomly generated initial controls (see Eq. (\ref{eq:random-controls})) to identify optimal straight {control} shots {{along with} monotonic climbs of $\mathcal{P}_{i\to f}$}, with numerical setups and implementation details discussed in Section \ref{sec:implementations}. For $1 \leq k \leq 10,000$, we use $\epsilon^k(s', s, t)$ to denote the controls involved in the $k^\text{th}$ run. More specifically, $\epsilon^k(s', 0, t), 0 \leq s' \leq S'$ represents the D-MORPH trajectory of the bottom controls, and $\epsilon^k(S', s, t), 0 \leq s \leq S(S')$ represents the final optimal {monotonic climb}, with bottom and top controls denoted by $\epsilon^k(S', 0, t)$ and $\epsilon^k(S', S(S'), t)$, respectively.

STSA successfully identified optimal straight shots for all 10,000 runs. However, 33 out of the 10,000 runs encountered non-monotonicity of their fidelity {climbs} $J(s', S(s'))$ due to discontinuity of $S(s')$ at some $s'$. {This} behavior can lead to numerical instability such as kinks {of} $J(s', S(s'))$, {and careful numerical calculations overcame those few encounters}. A topological analysis of $J(s', S(s'))$ and $S(s')$ is necessary to understand the origin of these {apparent} non-monotonicities, but this lies beyond the scope of this work. 
\begin{figure}[htbp]
\includegraphics[width=0.95\linewidth]{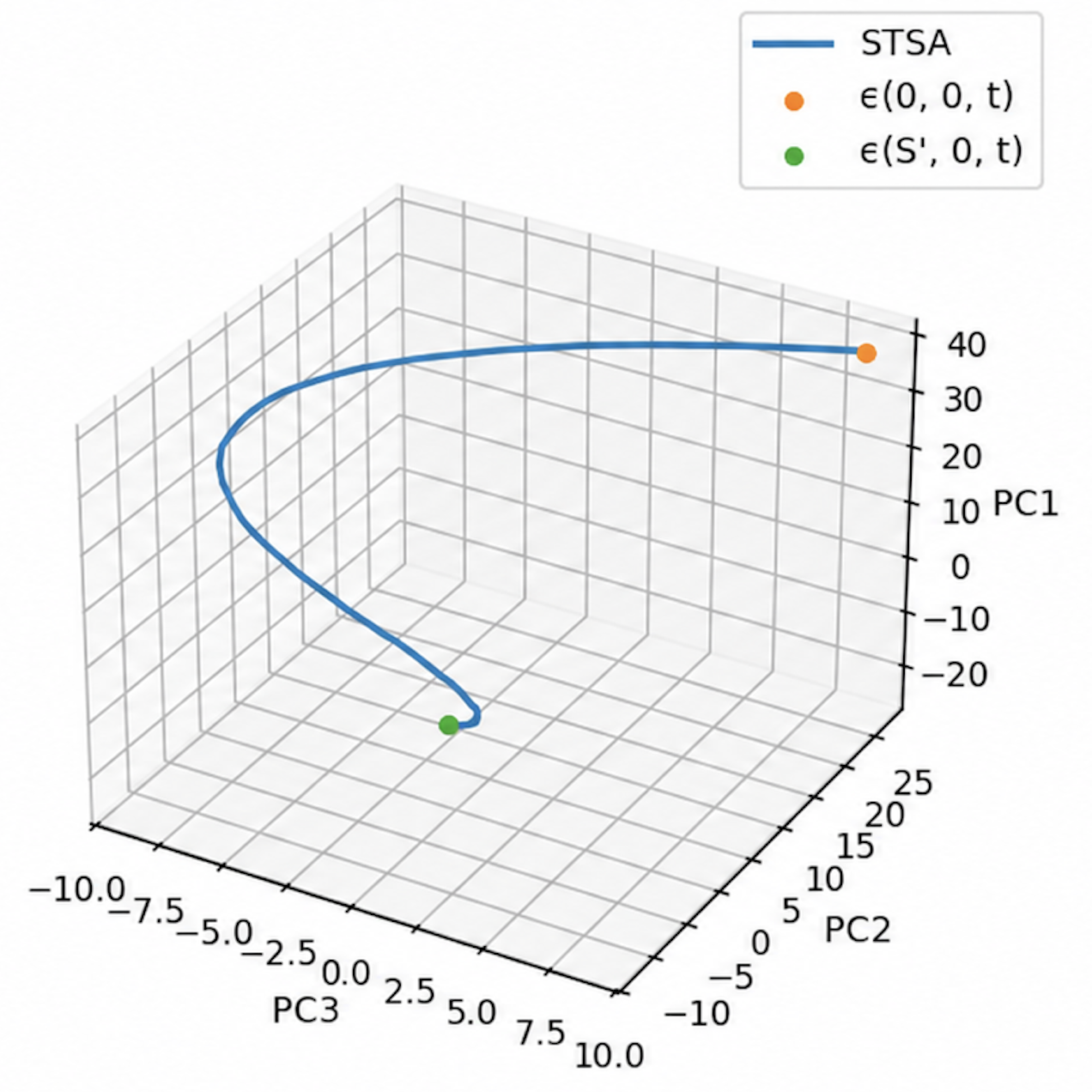}
    \caption{A dimensionality-reduced illustration of a bottom control trajectory produced by STSA. The controls are projected onto a three-dimensional subspace with {a} basis {consisting of} the top three principal components {PC1, PC2, PC3} of the entire control trajectory {explaining approximately {the total variance, respectively are 79\%, 19\%, and 2\%}}. Compared to the trajectory produced by random walk in Fig. \ref{fig:random-walk-trajectory}, the trajectory produced by STSA has an overall smoother shape and experiences {no} abrupt changes in direction.}
    \label{fig:trajectory-bottom}
\end{figure}
\begin{figure}[htbp]
    \includegraphics[width=0.95\linewidth]{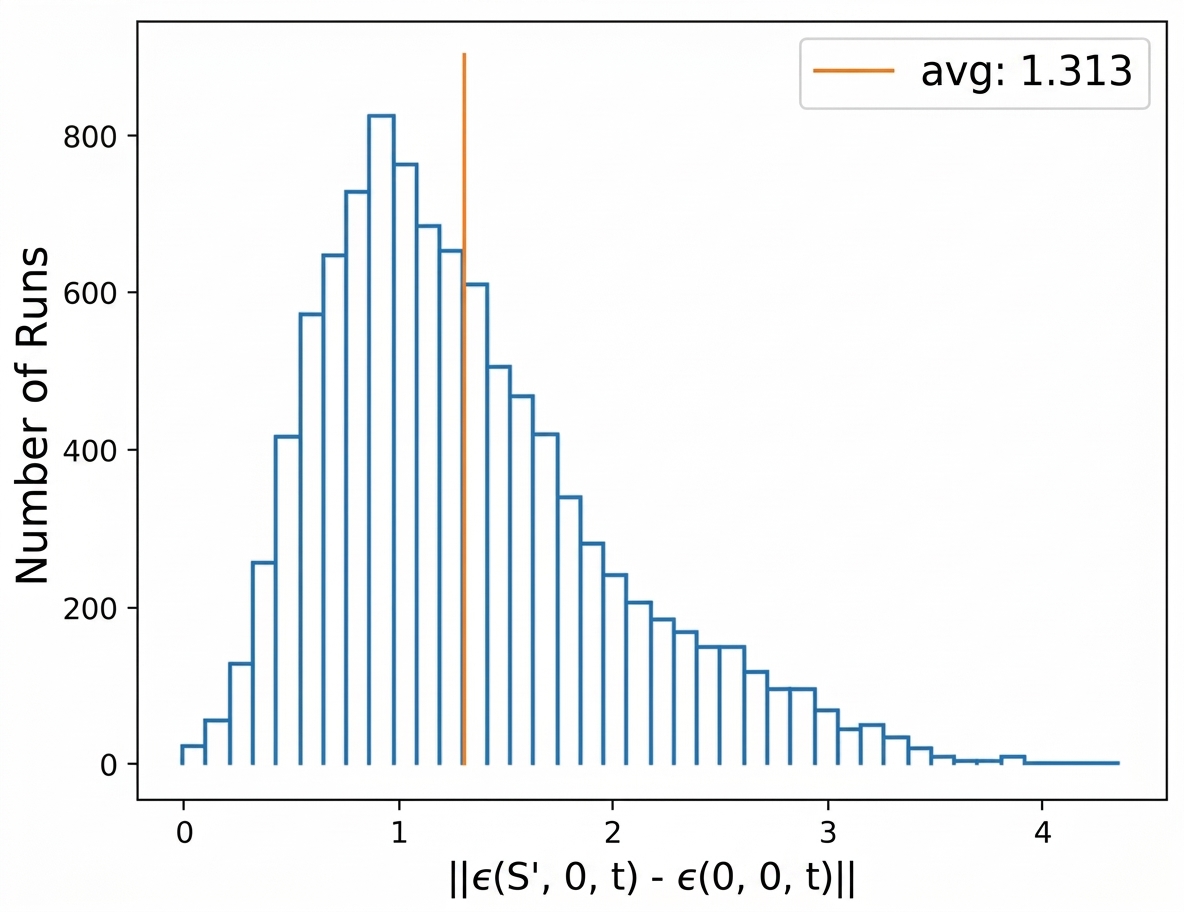}
    \caption{A histogram of the Euclidean distance traversed by the bottom controls over $0\leq s'\leq S'$ across 10,000 STSA runs. The average distance is approximately $1.31$, which is significantly {smaller} than the average distance of $2.93$ under random walk {in Fig. \ref{fig:distance-random-walk}} and relatively small compared to the average $L^2$-norm of the initial bottom controls of approximately $4.85$. }
    \label{fig:distance-bottom-control}
    \end{figure}

Besides the identification of optimal straight shots, we also examined their distribution in control space. In Fig. \ref{fig:trajectory-bottom}, we apply PCA to produce a dimensionality-reduced illustration of the bottom control trajectory for a representative STSA run. Compared to the trajectory produced by the random walk algorithm in Fig. \ref{fig:random-walk-trajectory}, the trajectory produced by STSA has an overall smooth shape and experiences {no} abrupt changes in direction.  We further compute the {end-to-end} Euclidean distances traversed by the bottom controls before reaching {the beginning of} an optimal straight shot, defined as ${\norm{\epsilon^k(S', 0, t) - \epsilon^k(0, 0, t)}_2}$, and Fig. \ref{fig:distance-bottom-control} presents a histogram of this metric across all runs. Compared to the Euclidean distances traversed under the random walk algorithm in Fig. \ref{fig:distance-random-walk}, the bottom controls identified by STSA traverse significantly shorter distances, with an average of $1.31$. {It was also found that the path length in the control space traversed by going from $s'=0$ to $S'$ on the $J=0.01$ level set is much smaller using STSA than random walk, which is evident in comparing the typical cases in Figs. \ref{fig:random-walk-trajectory} and \ref{fig:trajectory-bottom}.} Moreover, with the average $L^2$-norm of the initial bottom controls at $s' = 0$ being approximately $4.85$, the bottom controls do not need to {travel} far to find an optimal straight shot using STSA, further demonstrating the high prevalence of optimal straight shots in control space.

\begin{figure}[htbp]

        \includegraphics[width=0.95\linewidth]{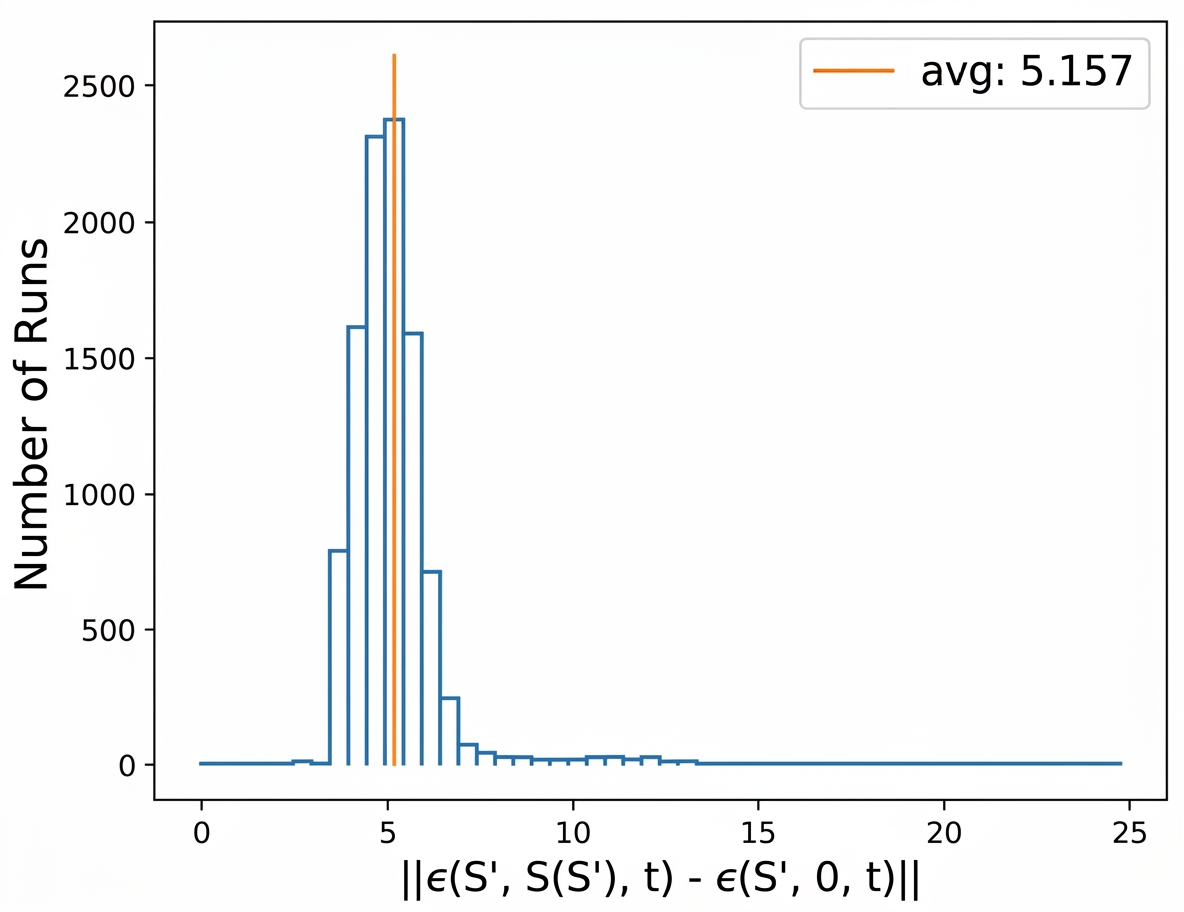}
    \caption{A histogram of the lengths of the optimal straight shots across 10,000 STSA runs, with an average of approximately 5.15.}
    \label{fig:st-length}
\end{figure}

To qualitatively investigate straight-shot climbs after an optimal straight shot has been identified, we compute the lengths of the optimal straight shots, defined as $\norm{\epsilon^k(S', S(S'), t)- \epsilon^k(S', 0, t)}_2$. A histogram of the lengths across all 10,000 runs is provided in Fig. \ref{fig:st-length}, with an average of approximately $5.15$. Once an optimal bottom control is identified, {the corresponding top optimal control can be readily obtained through the monotonic climb.}

\begin{figure}[htbp]
\subfloat[Nearest-neighbor distances of the bottom {initiating} controls.
]{
    \includegraphics[width=0.47\textwidth]{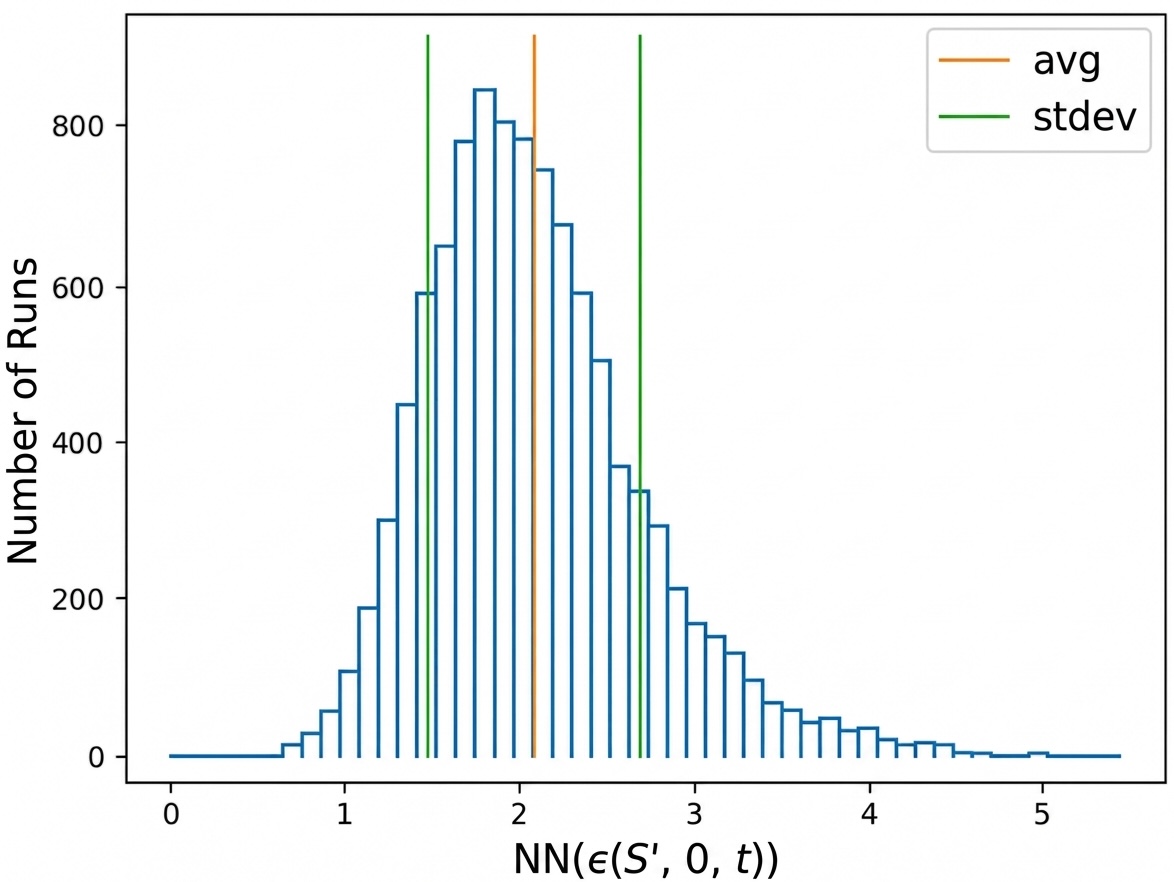}
    \label{fig:NN-bottom}
}
\hfill
\subfloat[Nearest-neighbor distances of the top {final} controls
]{
    \includegraphics[width=0.48\textwidth]{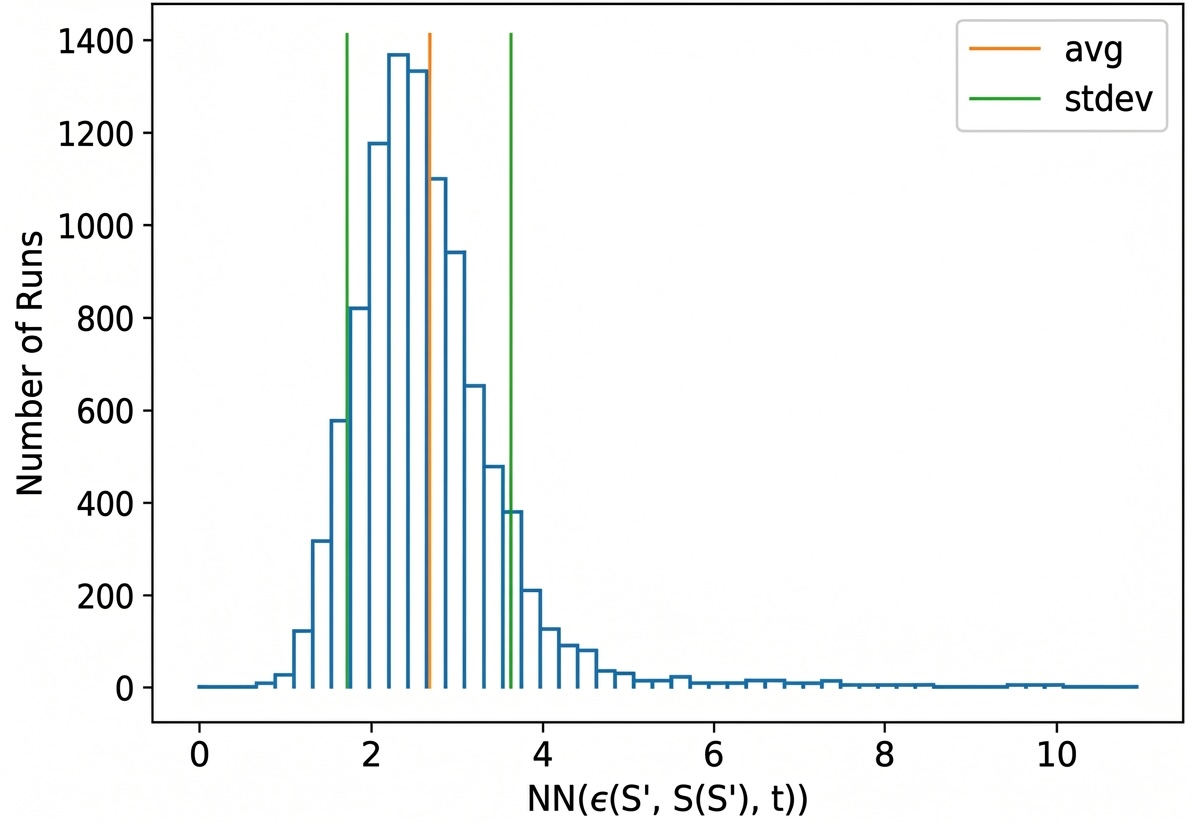}
    \label{fig:NN-top}
}
    \caption{(a)  Nearest-neighbor distances of the bottom {initiating} controls across 10,000 optimal straight shots. 
    The histogram appears approximately Gaussian, with an average of $2.06$ and a standard deviation of $0.59$. (b) Nearest-neighbor distances of the top {final} controls across 10,000 optimal straight shots. 
    The histogram appears approximately Gaussian, with an average of $2.65$ and a standard deviation of $0.94$. Both histograms exhibit approximately Gaussian shapes, but the one for the top controls has a larger variance and shows right-side outliers, possibly corresponding to runs with {pairs of }initial gradients that differ substantially in direction from {each other}.}
    \label{fig:NN}
\end{figure}

\begin{figure}[htbp]
    \centering    \includegraphics[width=0.95\linewidth]{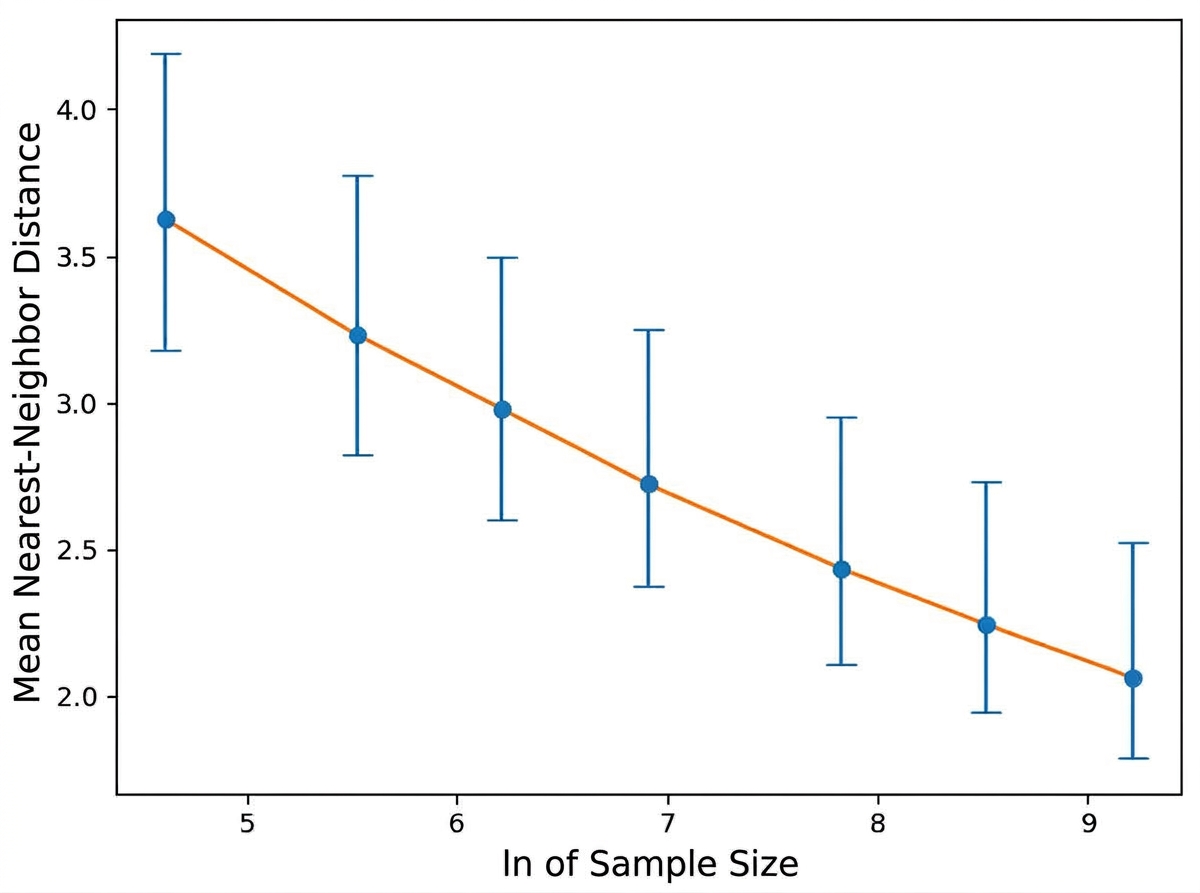}
    \caption{Mean nearest-neighbor distance of the bottom controls in the optimal straight-shot samples, {each at its respective $S'$ value,} versus the logarithm of the sample size. As the sample size increases, the mean nearest-neighbor distance decreases, potentially converging to zero; {see the text for further discussion}.}
    \label{fig:NN-size}
\end{figure}

We also adopt a nearest-neighbor distance metric to characterize the density of the bottom and top controls \cite{clark1954distance}. Given a set of controls, the nearest-neighbor distance of a control, denoted by $\mathsf{NN}(\epsilon(t))$, is defined as the smallest distance to its neighbors. For instance, given a bottom {starting} control {of a straight shot} $\epsilon^k(S', 0, t)$,
\begin{align}
~\label{eq:nnd}
\textsf{NN}(\epsilon^k(S',0,t)) = \min_{k'\neq k} \norm{\epsilon^{k'}(S',0,t) - \epsilon^k(S',0,t)}_2.
\end{align}
For the optimal straight shots identified by STSA, we present histograms of the nearest-neighbor distances for both the bottom and top controls {(i.e., corresponding to the initial and final controls of straight shots)} in Fig. \ref{fig:NN}. The overall Gaussian shapes observed in both histograms suggest relatively uniform distributions of the top and bottom controls, possibly inherited from the randomly sampled {trial} controls from Eq. (\ref{eq:random-controls}). The nearest-neighbor distances of the top controls in Fig. \ref{fig:NN-top} exhibit slightly greater variance, which can be explained by the climb algorithm spreading out controls with different initial gradients. 
To investigate how the sample size affects the density of optimal straight shots, we compute the average of the nearest-neighbor distances of the bottom controls, {each at their respective $S'$ value,} over randomly selected subsets of runs, with sample sizes ranging from 100 to 10,000 (i.e., 100, 250, 500, 1000, 2500, 5000, and 10,000), and a semi-log plot is made in Fig. \ref{fig:NN-size}. We conjecture, based on the overall converging trend in Fig. \ref{fig:NN-size}, that the bottom controls {consisting} of optimal {initial} straight-shot {fields} are not only prevalent but {appear to} form connected neighborhoods on the bottom level set, which will be explored in future work. {In considering this analysis, the clustering of all initial controls from Eq. (\ref{eq:random-controls}) means that these controls are constrained in the definition of the parameters in Eq. (\ref{eq:random-controls}). However, after the initial choice of control from Eq. (\ref{eq:random-controls}), no further such constraints enter into the execution of STSA.}
\section{Concluding remarks}
Three sufficient and necessary conditions {were found to specify straight shots through control space. These conditions are (i) the initial control on the bottom {level set} is connected to the final control on the top {level set} by a straight path through control space, (ii) the final control is optimal in that the objective reaches the top {level set}, and (iii) the climb of the {fidelity} landscape from the bottom to the top is monotonic} (see Section \ref{sec:intro}). Utilizing D-MORPH, {the STSA algorithm} is developed to systematically locate optimal straight shots by identifying their bottom controls on a predefined bottom level set, achieving significantly lower computational cost than {the prior} stochastic algorithm in \cite{PhysRevA.88.033425} and the random walk algorithm discussed in Section \ref{sec:random-walk-alg}. With tens of thousands of optimal straight shots generated using STSA {(i.e., including all three classes of physical control objectives)}, statistical analysis is used to examine their distribution in control space, identifying structured patterns such as normally distributed straight-shot lengths and decreased nearest-neighbor distances over increasing sample sizes. A theoretical {and/or computational} justification remains for {the few} cases in which $J(s', S(s'))$ is non-differentiable with respect to $s'$, which may prevent STSA from finding optimal straight shots. A topological study of $S(s')$ is necessary to explain STSA's behavior near {such} discontinuities. We emphasize that, {although STSA identifies optimal straight shots in a non-stochastic manner, it} remains significantly more expensive than steepest ascent alone {with no attempt at finding straight control shots}, and therefore {STSA in its present form} does not accelerate the {existing} optimal control algorithms. {The prevalence of optimal straight shots, however, suggests the possibility of developing more efficient optimal control algorithms, which is left for future studies.} {A fundamental distinction between straight shot landscape climbs versus the like climbs with the standard gradient algorithm is that the former are relaxed to only be monotonic, whereas the latter is steepest ascent. Understanding this distinction at a fundamental mathematical level could aid in finding the aforementioned {route} to a high-efficiency straight shot algorithm.}

While only the state-to-state transition probability problem was discussed in the results for illustrative purposes {between level sets at $0.01$ and $0.99$}, {these numerical results were also extended to the top and bottom level sets being $0.001$ and $0.999$, respectively. The reported qualitative findings remained the same.} We also verified that STSA successfully identified optimal straight shots for the $\Tr(\rho O)$ and unitary {transformation generation} problems defined in Eqs. (\ref{eq:tr(rhoO)}) and (\ref{eq:U}); however, due caution is required when selecting appropriate $\Delta s$ and $\Delta s'$ to maintain high numerical accuracy. Since the definition of a straight shot is highly generic, we also expect STSA to be applicable for identifying straight-shot solutions for {classical mechanical} optimization problems {where landscape traps {appear} not {to be} present} \cite{joe2013topology,joe2015topology}. {This circumstance may apply to other dynamical control problems as well \cite{russell2018control, russell2017common}.}

All the numerical evidence supports a high prevalence of efficient optimization pathways (optimal straight shots) in {quantum} control space. We further conjecture that there are infinitely many such straight shots, and their bottom controls likely form connected neighborhoods; identifying them will likely require more sophisticated landscape exploration {analysis and} algorithms, which we leave to future work.

\section{Acknowledgment}
The authors thank Amr Sobeh for helpful advice on graphics. {The authors Y.C. and H.R. acknowledge {partial} support from the Department of Energy [Basic Energy Sciences] under Award Number(s) [DE-FG02-02ER15344].
This report was prepared as an account of work sponsored by an agency of the United States Government. Neither the United States Government nor any agency thereof, nor any of their employees, makes any warranty, express or implied, or assumes any legal liability or responsibility for the accuracy, completeness, or usefulness of any information, apparatus, product, or process disclosed, or represents that its use would not infringe privately owned rights. Reference herein to any specific commercial product, process, or service by trade name, trademark, manufacturer, or otherwise does not necessarily constitute or imply its endorsement, recommendation, or favoring by the United States Government or any agency thereof. The views and opinions of authors expressed herein do not necessarily state or reflect those of the United States Government or any agency thereof.}
\section{Data Availability}
The numerical data supporting the findings of this study and the corresponding computational code are available from the authors upon reasonable request.
\appendix
\section{Derivation of Eq. (\ref{eq:dj/ds'>0})}\label{appendix1}
We provide the intermediate derivations for Eq. (\ref{eq:dj/ds'>0}). By the chain rule, we have
\begin{widetext} 
\begin{align}
\frac{\partial J(s', S(s'))}{\partial s'} =\label{eq:appendix1}&\left[\frac{\partial J(s', s)}{\partial s'}\right]_{s=S(s')} +  \left[\frac{\partial J(s', s)}{\partial s}\right]_{s=S(s')}\frac{\partial S(s')}{\partial s'} \intertext{Given the optimality condition defined in Eq. (\ref{eq:optimality}), $\left[\frac{\partial J(s', s)}{\partial s}\right]_{s=S(s')}=0$ for all straight shots, the second term in Eq. (\ref{eq:appendix1}) vanishes. }\\\frac{\partial J(s', S(s'))}{\partial s'} =&\left[\frac{\partial J(s', s)}{\partial s'}\right]_{s=S(s')} \\
=&\left[\int_0^T dt\frac{\delta J(s', s)}{\delta \epsilon(s', s, t)}\frac{\partial \epsilon(s', s,t)}{\partial s'}\right]_{s=S(s')}\\
=&\left[\int_0^T dt \frac{\delta J(s', s)}{\delta \epsilon(s', s, t)}\int_0^T dt'\left[\frac{\delta \epsilon(s', s,t)}{\delta \epsilon(s',0, t')}\right]_{s=0}\frac{\partial \epsilon(s',0,t')}{\partial s'}\right]_{s=S(s')}
\intertext{
By the straightness condition defined in Eq. (\ref{eq:straight2}) {including now the implicitly present variable $s'$}, we have}
=&\left[\int_0^T dt \frac{\delta J(s', s)}{\delta \epsilon(s', s, t)}\int_0^T dt'\left(\delta(t-t') + s\left[\frac{\delta^2 J(s',s)}{\delta \epsilon(s',s,t') \delta \epsilon(s',s,t)}\right]_{s=0}\right) \frac{\partial \epsilon(s',0,t')}{\partial s'}\right]_{s=S(s')}\\
=&\left[\int_0^T dt \frac{\delta J(s', s)}{\delta \epsilon(s', s, t)}\frac{\partial \epsilon(s',0,t)}{\partial s'} + s\int_0^T dt\int_0^T dt'\left[\frac{\delta^2 J(s',s)}{\delta \epsilon(s',s,t') \delta \epsilon(s',s,t)}\right]_{s=0}\frac{\delta J(s', s)}{\delta \epsilon(s', s, t)}\frac{\partial \epsilon(s',0,t')}{\partial s'}\right]_{s=S(s')}
\intertext{
{As $s$ is in the brackets, we have}}
=&\int_0^T dt \left(\left[\frac{\delta J(s', s)}{\delta \epsilon(s', s, t)}\right]_{s=S(s')} +S(s')\int_0^T dt'\left[\frac{\delta^2 J(s',s)}{\delta \epsilon(s',s,t) \delta \epsilon(s',s,t')}\right]_{s=0}\left[\frac{\delta J(s', s)}{\delta \epsilon(s', s, t')} \right]_{s=S(s')}\right)\frac{\partial \epsilon(s',0,t)}{\partial s'}.
\end{align}
\end{widetext}
\bibliography{apssamp}
\end{document}